\documentclass[aps,prd,nofootinbib,twocolumn,amsmath,amssymb,floatfix,floats,titlepage,superscriptaddress]{revtex4-1}
\usepackage{bm}
\usepackage{amsmath}
\usepackage{epsfig}
\usepackage{color}
\usepackage{natbib}
\usepackage{graphicx,epsfig}
\usepackage{hyperref}
\usepackage{ifthen}
\usepackage{xstring}
\usepackage{graphicx}

\usepackage[applemac]{inputenc}
\usepackage{amsmath,amssymb} 
\usepackage{color}
\usepackage{epstopdf}

\AppendGraphicsExtensions{.gif}

\newcommand{\physrep}{Physics Reports}
\newcommand{\aapr}{A\&A Rev.}
\newcommand{\apjl}{ApJL}

\newcommand{\aap}{A\&A}
\newcommand{\mnras}{MNRAS}

\newcommand{\jcap}{{\em JCAP }}
\newcommand{\aj}{AJ }

\def\be{\begin{equation}}
\def\ee{\end{equation}}
\def\ben{\begin{equation*}}
\def\een{\end{equation*}}

\def\ba{\begin{eqnarray}}
\def\ea{\end{eqnarray}}
\def\ban{\begin{eqnarray*}}
\def\ean{\end{eqnarray*}}

\newcommand{\refsec}[1]{\S\ref{sec:#1}}
\newcommand{\refeq}[1]{Eq.~(\ref{eqn:#1})}

\newcommand{\reffig}[1]{Fig.~\ref{fig:#1}}

\newcommand{\reftab}[1]{Tab.~\ref{tab:#1}}

\definecolor{darkgreen}{cmyk}{0.85,0.2,1.00,0.2} 
\definecolor{purple}{cmyk}{0.5,1.0,0,0}

\usepackage{verbatimbox}
\usepackage{float}
\usepackage{subfig}
\usepackage{hyperref}
\usepackage{printlen}
\newcommand{\begm}{\begin{pmatrix}}
\newcommand{\enm}{\end{pmatrix}}

\begin{document}
\draft
\title{Mapping the Integrated Sachs-Wolfe Effect}

\author{A. Manzotti}
\affiliation{Department of Astronomy \& Astrophysics, University of Chicago, Chicago IL 60637}
\affiliation{Kavli Institute for Cosmological Physics, Enrico Fermi Institute, University of Chicago, Chicago, IL 60637}
\author{S. Dodelson}
\affiliation{Fermilab Center for Particle Astrophysics, Fermi National Accelerator Laboratory, Batavia, IL 60510-0500}
\affiliation{Department of Astronomy \& Astrophysics, University of Chicago, Chicago IL 60637}
\affiliation{Kavli Institute for Cosmological Physics, Enrico Fermi Institute, University of Chicago, Chicago, IL 60637}

\date{\today}
\begin{abstract}

On large scales, the anisotropies in the cosmic microwave background (CMB) reflect not only the primordial density field but also the energy gain when photons traverse decaying gravitational potentials of large scales structure, what is called the Integrated Sachs-Wolfe (ISW) effect. Decomposing the anisotropy signal into a primordial piece and an ISW component, the main secondary effect on large scales, is more urgent than ever as cosmologists strive to understand the Universe on those scales. We present a likelihood technique for extracting the ISW signal combining measurements of the CMB, the distribution of galaxies, and maps of gravitational lensing. We test this technique with simulated data showing that we can successfully reconstruct the ISW map using all the datasets together. Then we present the ISW map obtained from a combination of real data: the NVSS galaxy survey, temperature anisotropies and lensing maps made by the Planck satellite. 
This map shows that, with the datasets used and assuming linear physics, there is no evidence, from the reconstructed ISW signal in the Cold Spot region, for an entirely ISW origin of this large scale anomaly in the CMB.
However a large scale structure origin from low redshift voids outside the NVSS redshift range is still possible. Finally we show that future surveys, thanks to a better large scale lensing reconstruction will be able to improve the reconstruction signal to noise which is now mainly coming from galaxy surveys.
\end{abstract}

\maketitle

\section{Introduction}\label{sec:intro}
One of the lingering cosmological mysteries is the structure of the Universe on the largest observable scales. There are maddeningly few handles on this ultra-large scale structure, and the primary observational source of information -- anisotropies in the cosmic microwave background (CMB) -- has produced as much confusion as clarity. Indeed the power spectrum of the temperature anisotropies appears to be lower than expected on the largest scales, and the moments are aligned with a so-called ``Axis of Evil'' \cite{de-oliveira-costa:2004}. Additionally some regions of the sky are colder than expected \cite{vielva:2004} and an hemispherical asymmetry in the large scale power spectrum \cite{eriksen:2007} seems to be present. Furthermore the recent BICEP2 results \cite{ade:2014} point to another potential anomaly: the possibility that gravitational waves are produced in the early universe but do not leave a clear signature in the temperature anisotropy. One can even speculate that the observed acceleration of the Universe is also a very large scale effect so might be related to these other anomalies. 

For each of these, solutions have been proposed, but cosmic variance limits the number of measurable modes so makes it particularly difficult to determine whether these anomalies are real or simply statistical fluctuations, and if the former, to pinpoint the underlying physical cause. One possible way to make headway is to decompose the large scale anisotropy field into a primordial part that reflects the conditions in the Universe at very early times and a late-time component due to secondary effects caused by the interaction of CMB photons with the large scale structures of the universe. If we can achieve this goal, we can at least reduce the set of possible explanations: if an anomaly can be attributed to late-time effects, then primordial explanations can be eliminated. 

On these very large scales, the main secondary effect is the late Integrated Sachs-Wolfe (ISW) effect \cite{sachs:1967}.
Climbing in gravitational potentials, the CMB photons get blue-shifted and, if potentials are static as expected in a matter dominated universe they get red-shifted by the same amount when they climb out, resulting in a zero net effect.
However at low redshift, during the accelerated expansion of the universe, the gravitational potentials decay with time, changing the energy of the photons. 

Here we describe a way of extracting a map of the ISW signal -- i.e., the part of the CMB temperature anisotropy map that is due to the ISW effect -- that uses information not just from the CMB, but also from external maps of large scale structure~\cite{dupe:2011,nolta:2004,xia:2009,scranton:2003,crittenden:1996,pietrobon:2006,mcewen:2008,raccanelli:2008,giannantonio:2008,pietrobon:2006,vielva:2006,afshordi:2004,giannantonio:2012,zhang:2006,francis:2010,Barreiro:2008sn,2009ApJ...701..414G,barreiro:2013,planck-ISW,rassat:2013,2008MNRAS.391.1315F,frommert:2008,schiavon:2012,fosalba:2003,fosalba:2004,boughn:2004}. 
Indeed maps of the distribution of galaxies and maps of gravitational lensing are in principle correlated with the ISW signal since it is precisely the over-dense regions in these maps that produce the ISW effect. We propose an optimal way to flexibly combine all this information and extract the desired signal adopting a likelihood approach. \refsec{estimator} describes this method; \refsec{theory} presents the calculations needed to implement it; \refsec{data} the data sets used. \refsec{results} presents our main results, first testing our technique on simulated maps, then applying it to 3 current data sets -- the Planck temperature map; the Planck map of gravitational lensing obtained from the quadratic estimator of small scale anisotropies; and a map of radio galaxies measured in the NVSS survey -- and finally showing projections for upcoming surveys. Conclusions are given in \refsec{conclusions}. 

\newcommand\tisw{T^{\rm{ISW}}}
\newcommand\aisw{a^{\rm{ISW}}}
\section{Optimal map estimator} \label{sec:estimator}

The observed CMB temperature anisotropy can be decomposed into
\be
T^{\text{obs}}=T^{\text{p}}+T^{\rm{ISW}}+T^{\text{n}},
\ee
where $T^{\text{p}}$ is the primordial CMB generated at the last scattering surface; $T^{\rm{ISW}}$ is the late-time integrated Sachs-Wolfe effect uncorrelated with the primordial signal; and $T^{\text{n}}$ represents instrumental and atmospheric noise.
Each of these three components is assumed to be uncorrelated with the others and  drawn from a Gaussian distributions with mean zero and a known covariance matrix. Actually some residual correlation  between the ISW and the primordial signal is present but the amplitude is less than a percent of the ISW power at all scales and, for this reason, we neglect it.

Therefore, the combination $T^{\text{obs}}-T^{\rm{ISW}}=T^{\text{p}}+T^{\text{n}}$ will be distributed as a Gaussian with covariance matrix $C\equiv C^{\text{p}}+C^{\text{n}}$, i.e., equal to the sum of the covariance matrices of the noise and the primordial CMB.

So, the likelihood of obtaining CMB data as a functional of the ISW contribution is:

\small
\ba \label{eqn:likelihood1}
\mathcal{L}\left(\tisw\right) &\propto& \frac{1}{\sqrt{\det(C)}}\nonumber \\
&\times& \exp\left\{ -\frac{1}{2}\left(T^{\text{obs}}-T^{\rm{ISW}}\right)\,C^{-1}\,\left(T^{\text{obs}}-T^{\rm{ISW}}\right)
\right\} .\nonumber \\
\ea
\normalsize

This is still shorthand since the temperature is measured at many different points on the sky. Throughout, we will decompose the fields into spherical harmonics, so that 
\be
\tisw(\vec\theta) = \sum_{lm} a^{\rm ISW}_{lm} Y_{lm}(\vec\theta).
\ee
We are also assuming that the covariance matrix for the $a_{lm}$'s depends only on $l$. 
 
If the CMB temperature were the only data set we were using, we could still improve on this likelihood by including a prior on $\tisw$: it too is drawn from a Gaussian distribution with mean zero and a covariance matrix that (as we will see in the next section) is straightforward to compute. If other data sets that trace the large scale structure were available, these could be included in the likelihood as well. Indeed $\tisw$ is correlated with these data sets, so forming a vector $\vec d$ that contains first $\tisw$ and then the various tracers, the combined likelihood would be the product

\ba \label{eqn:likelihood2}
\mathcal{L} &\propto& \det(CD)^{-1/2} \exp \left\{ -\frac{1}{2}d^{t} D^{-1}d \right\} \nonumber \\ 
&\times& \exp\left\{ -\frac{1}{2}\left(T^{\text{obs}}-T^{\rm{ISW}}\right)\,C^{-1}\,\left(T^{\text{obs}}-T^{\rm{ISW}}\right)
\right\},\nonumber \\
\ea

where $D$ is the covariance matrix of the vector $\vec d$. To be concrete, suppose $\vec d$ includes two tracers of large scale structure, a map of galaxy over-densities $\delta^g$, and a map of the projected gravitational potential $\delta^\phi$, so that

\be\label{eqn:dvector}
\vec d_{\ell m} \equiv (\aisw_{\ell m},a^g_{\ell m},a^\phi_{lm})
\ee
where here the decomposition into spherical harmonics is explicit.
The covariance matrix in this 3D case is
 \be
 D_l =\left(\begin{matrix}
    C_{\ell}^{\rm{ISW},\rm{ISW}} & C_{\ell}^{g,\rm{ISW}}& C_{\ell}^{\phi,\rm{ISW}}  \cr
 C_{\ell}^{g,\rm{ISW}} & C_{\ell}^{gg}   & C_{\ell}^{g,\phi} \cr
C_{\ell}^{\phi,\rm{ISW}} & C_{\ell}^{\phi,g}   & C_{\ell}^{\phi\phi} 
 \end{matrix} \right).
 \label{eqn:Dmatrix}
\ee
This gives us the first line of \refeq{likelihood2}.
Similarly the argument of the exponential on the second line of \refeq{likelihood2} would contain $\aisw_{lm}$ as well as $a^{\rm obs}_{lm}$ and the appropriate covariance matrix $C_l$, so the likelihood has an $\aisw_{lm}$ dependence in both terms. 
The optimal $\aisw_{\ell m}$ estimator is obtained by minimizing the likelihood in \refeq{likelihood2}:
\be 
\hat{a}^{\rm ISW}_{\ell m}=N_{l}\left[C_{\ell}^{-1} a_{\ell m}^{\text{obs}} - (D^{-1}_{\ell})_{12}~a^g_{\ell m} - (D^{-1}_{\ell})_{13}~a^\phi_{\ell m}\right]
\ee
where
\be\label{eqn:error}
N_{\ell}^{-1}=C_{\ell}^{-1}+(D^{-1}_{\ell})_{11}
\ee
estimates the reconstruction variance, being the second derivative of the log of the likelihood. This estimator can be easily extended.
If there are $n$ tracers of large scale structure, such as galaxies in different redshift slices and maps from gravitational lensing, then the estimator generalizes to
\be\label{eqn:estimator}
\hat{a}^{\rm ISW}_{\ell m}=N_{l}\left[C_{\ell}^{-1} a_{\ell m}^{\text{obs}} - \sum_{j=2}^{n+1} (D^{-1}_{\ell})_{1j}~a^j_{\ell m}\right]
\ee
with the variance still given by \refeq{error}.

Note that \refeq{estimator} reduces to ISW estimators that can be found in the literature for specific subsets of external tracers.
For example, when $\vec d_{\ell m}$ is a 1D vector with $\aisw$ only, the estimator is the usual Wiener filter estimator $\hat a_{\ell m}^{\rm{ISW}}=\frac{C^{ISW}}{C^{p}+C^{n}}a_{\ell m}^{\text{obs}}$. Using only one galaxy overdensity bin we find the estimator $\hat a_{\ell m}^{\rm{ISW}}=\frac{C^{g-\rm{ISW}}}{C^{g}}a_{\ell m}^{\text{obs}}$ used for example in \cite{francis:2010}, and using alternatively lensing information or one galaxy bin together with CMB temperature data we recover the estimator proposed \cite{Barreiro:2008sn} and successfully used to reconstruct the ISW map in \cite{barreiro:2013} and \cite{planck-ISW}.


\section{Theoretical covariances}  \label{sec:theory}

The fiducial gaussian covariances in \refeq{estimator} can be computed theoretically. Here we present the general calculation for the covariance in multipole space of two 2D fields. A random field defined on the sky $\delta(\mathbf{\hat n})$ includes information from the full 3D field $\delta(\vec x,t)$ 
integrated along the line of sight:
\be \label{eqn:lineofsightint}
\delta^X(\mathbf{\hat n})=\int \text{d}\chi ~\delta\left(\vec x[\chi,\hat n],t(\chi)\right)~W^X(\chi),
\ee
where $\chi$ is the comoving distance along the line of sight; $\delta$ is the 3D matter over-density; and $W^X(\chi)$ is a weighting-window function specific to the field $X$. Along the line of sight, $\delta$ is evaluated at 3D position $\vec x$ and time $t$ determined by the radial position $\chi$ and angular position $\hat n$. In the small sky approximation, for example, $\vec x = [\chi\hat n,\chi]$, but we will work in the full sky expressions because most of the signal of interest is on large scales.  
We will consider three sets of 2D fields in this paper: the density of galaxies in different redshift slices, the ISW signal itself, and the projected gravitational potential along the line of sight inferred from maps of lensing. Let us focus on the window function for each of these fields.

The window function for galaxies can be simplified by the assumptions (for example in the context of the peak-background split model \cite{cole:1989,bardeen:1986}) that on large scales the bias between the galaxy density and matter density is independent of scale. Therefore, we set
\be\label{eqn:gal-kernel}
W^{g_i}(\chi) = b_i \frac{dN_i}{d\chi}
\ee
where $b_i$ is the bias of the sample $i$ and $dN_i/d\chi$ is the differential number of galaxies in this slice as a function of $\chi$, normalized so that its integral over $\chi$ is one. The bias can be fixed given a background cosmology by using the auto-correlation function.

The ISW signal is sensitive to changes in the gravitational potential along the line of sight:
\begin{equation}\label{eqn:deltaTISW}
\delta_T^{\rm \rm{ISW}} (\mathbf{n})=2\int_0^{\infty} ~ \frac{d\chi}{1+z(\chi)}\, \textrm{e}^{-\tau(\chi)}\frac{\partial\Phi(\vec x(\chi,\hat n),t(\chi))}{\partial t}
\end{equation}
where $\tau(\chi)$ is the optical depth out to distance $\chi$. Practically, we can set $\tau$ to zero all the way back to recombination and then to infinity at larger distances. For perturbations sufficiently within the horizon, 
the potential can be related to the matter fluctuations $\delta$ in Fourier space via the Poisson equation
\begin{equation}\label{eqn:poisson}
\Phi({\bf k},z)=-\frac{3}{2}H_0^2\Omega_m (1+z)\frac{\delta({\bf k},z)}{k^2},
\end{equation}
where $H_0$ and $\Omega_m$ are respectively the today values of the Hubble constant and the ratio of the matter density to the critical density. On large scales the over densities evolve as $\delta(z)=\delta(z=0) D(z)$, where we normalize the growth function $D$ so that it is equal to one today.
Therefore, the window function depends on wavenumber as well as $\chi$:
\be
W^{ISW}(\chi,k) = -\Theta(\chi_*-\chi)\,\frac{3\Omega_m H_0^2}{k^2} \frac{\partial \ln([1+z]D(z))}{\partial t}
\ee
where the step function prefactor sets the window function to one all the way back to the last scattering surface at $\chi_*$.

Gravitational lensing probes the integrated gravitational potential back to the sources with a weighting factor:
\begin{eqnarray}
\delta^\phi(\mathbf{n})&=&- 2 \int_0^{\chi_s} d\chi
\frac{(\chi_s-\chi)}{\chi \chi_s}
\Phi (\vec x[\chi,\mathbf{n}];t(\chi)) \,.
\label{eqn:lenspotential}
\end{eqnarray}
Here we will focus on lensing maps made from the CMB (for a review see \cite{lewis:2006}), so that the sources are at $\chi_s=\chi_*$, the last scattering surface. Moving to Fourier space and converting the potential to the overdensity using the Poisson equation then leads to the lensing window function,
\be
W^\phi(\chi,k) = \Theta(\chi_*-\chi)\,\frac{3\Omega_m H_0^2 (1+z)}{k^2}\frac{(\chi_*-\chi)}{\chi\chi_*}.
\ee

Each of these fields $\delta^X(\mathbf{\hat n})$ on the sky can be expanded in spherical harmonics with coefficients given by
\be \label{eqn:sphericalexpansion}
a^X_{\ell m}=\int \frac{d\Omega}{4\pi}\, ~\delta^X(\mathbf{\hat n}) Y_{\ell m}^*(\mathbf{\hat n}).
\ee
To carry out the integral we Fourier transform the integrand of $\delta^X$ and then use the spherical harmonics plane wave expansion to arrive at
\be\label{eqn:dx}
a^X_{lm} =4\pi i^{\ell} \int\frac{d^{3}k}{(2\pi)^{3}} I^X_{\ell}(k)Y_{\ell m}^*(\mathbf{k})\, \delta(\vec k)
\ee
with 
\be
I^X_{\ell}(k)\equiv \int d\chi D(\chi)\, W^X(k,\chi)~  j_{\ell}(k\chi).\ee
Note that $\delta$ in \refeq{dx} is evaluated at the present, since the time dependence is governed by the growth function $D$ included in $I^{X}_{\ell}(k)$.

Finally the covariance $C^{XY}_{\ell}$ of two statistically isotropic gaussian fields is defined as $\langle a^{X}_{\ell m} a^{Y*}_{\ell' m'}\rangle=C^{XY}_{\ell}~\delta_{\ell\ell'}\delta_{mm'}$. Therefore,
\be \label{eqn:cldef}
C^{XY}_{\ell}=\frac{2}{\pi}\int_0^\infty k^{2} dk ~P(k) I^{X}_{\ell}(k)I^{Y}_{\ell}(k)
\ee
where $P(k)$ is the matter power spectrum today.


The covariances in the full covariance matrix $D$ from signal are now determined. The remaining issue is to include the noise in the measurements of the auto-spectra of $a_{lm}^g$ and $a_{lm}^\phi$. 
Because the main source of noise in estimating the galaxies density is Poisson, we need to modify \refeq{cldef} so that
\be\label{eqn:ggnoise}
C^{gg}_l \rightarrow C^{gg}_l + \frac{1}{n^{L}},
\ee
where $n_{L}$ is the surface density of sources per steradian.
Similarly, the auto-spectrum of gravitational lensing obtained from the CMB is modified using the expression for the reconstruction noise first obtained in Ref.~\cite{hu:2001}.

\section{Data Sets}\label{sec:data}

In order to reconstruct the ISW signal we need sky maps of the CMB temperature anisotropies and of large scale structure tracers as well as a fiducial cosmological model to compute the theoretical covariances derived in \refsec{theory}. 

\begin{figure}[tb]
\centering
  \includegraphics[width=\linewidth]{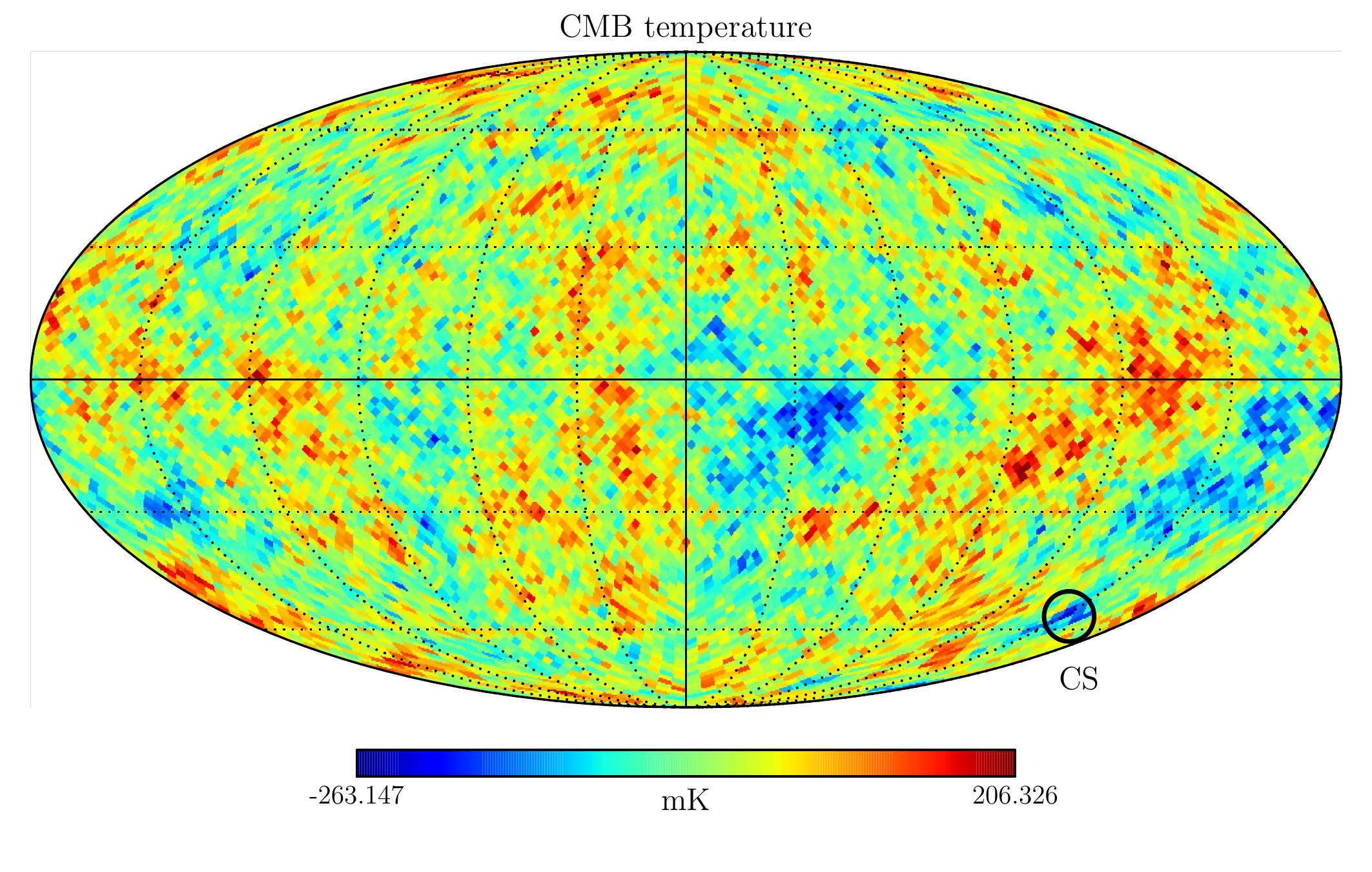}
  \caption{\footnotesize CMB Planck temperature map, $N_{\text{side}}=32$. The cold spot location is showed in the map.}
  \label{fig:cmbmap}
\end{figure}

For the CMB temperature, we use the SMICA map released by the Planck collaboration \cite{planck-collaboration:2013b}. This map is produced by linearly combining all the Planck input channels (from 30 to 857 GHz) with weights that vary with the multipole. At the angular scales we are considering, it is essentially limited by cosmic variance and foreground subtraction uncertainties. 
Because we are trying to reconstruct a large scale effect we do not need the high resolution the data are released in, so we downgrade the map using HEALPIX \cite{gorski:2005} to a resolution of of $N_{side}=32$ (\reffig{cmbmap}) which corresponds to 12288 pixels of equal area.

\begin{figure}[tb]
\centering
  \includegraphics[width=\linewidth]{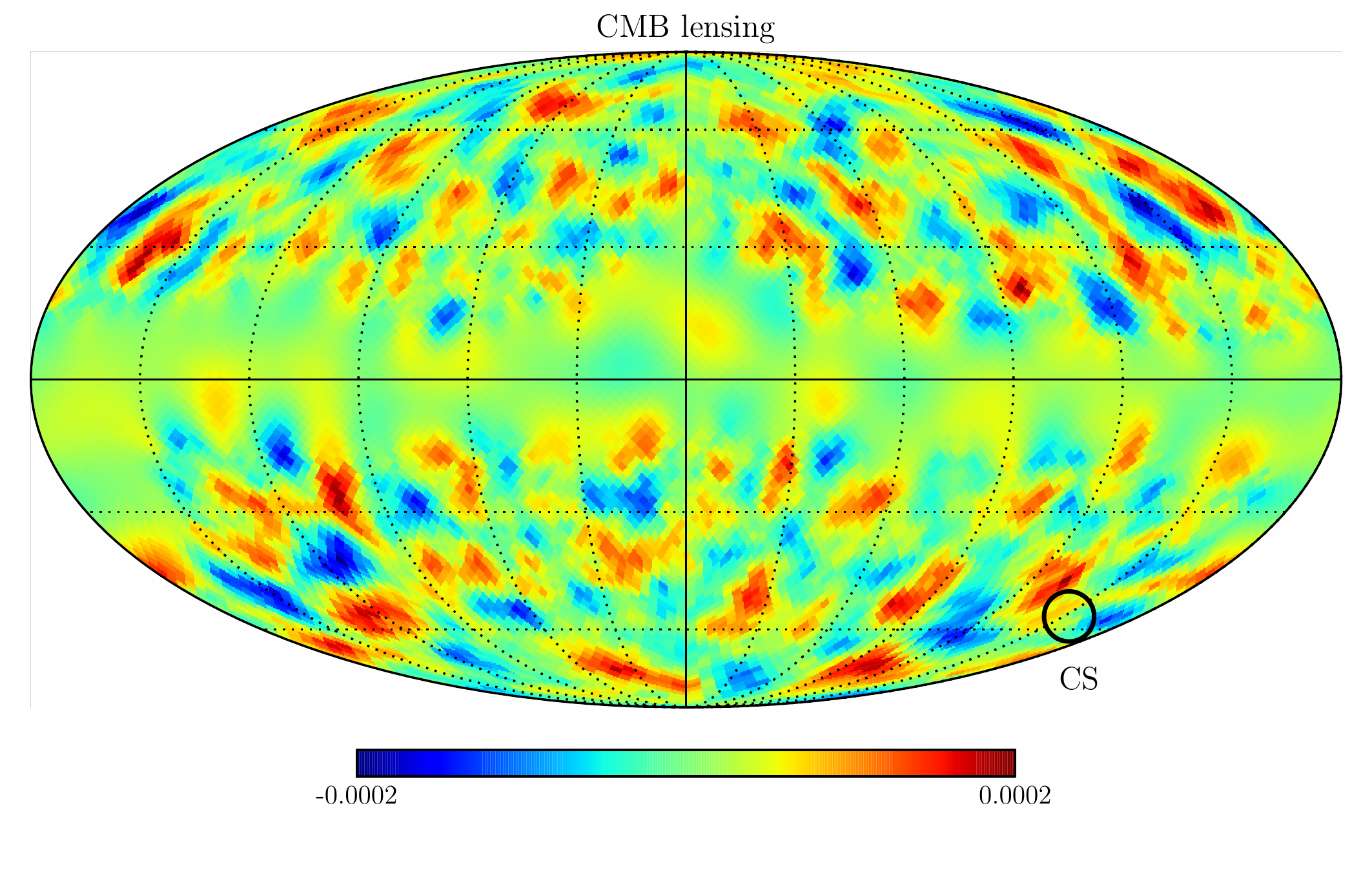}
  \caption{\footnotesize Planck gravitational lensing potential map, $N_{\text{side}}=32$. The cold spot location is showed in the map.}
  \label{fig:lensingmap}
\end{figure}

The second essential element of our reconstruction algorithm is a collection of sky maps of large scale structure tracers. We use the lensing potential reconstructed with CMB temperature information and a tracer of the matter density.
 
The lensing potential map together with the reconstruction noise was also distributed as a part of the Planck public data release \cite{planck-collaboration:2013b}. 
We refer to \cite{planck-collaboration:2013a} for a detailed description of the technique applied to reconstruct it from the temperature anisotropies of the CMB.
We normalize the distributed map of the un-normalized lensing potential estimator following \cite{planck-collaboration:2013a}. \reffig{lensingmap} shows the obtained map downgraded to the same resolution as the temperature map.

We use Luminous Active Galactic Nuclei as fair tracers of the matter density matter field.
These are very powerful sources and can be detected, for example in the radio band, out to high redshift. 
As a consequence they probe the large scale gravitational potential wells during the onset of the dark energy-driven accelerated expansion of the universe that generates the ISW effect.
\begin{figure}[tb]
\centering
  \includegraphics[width=\linewidth]{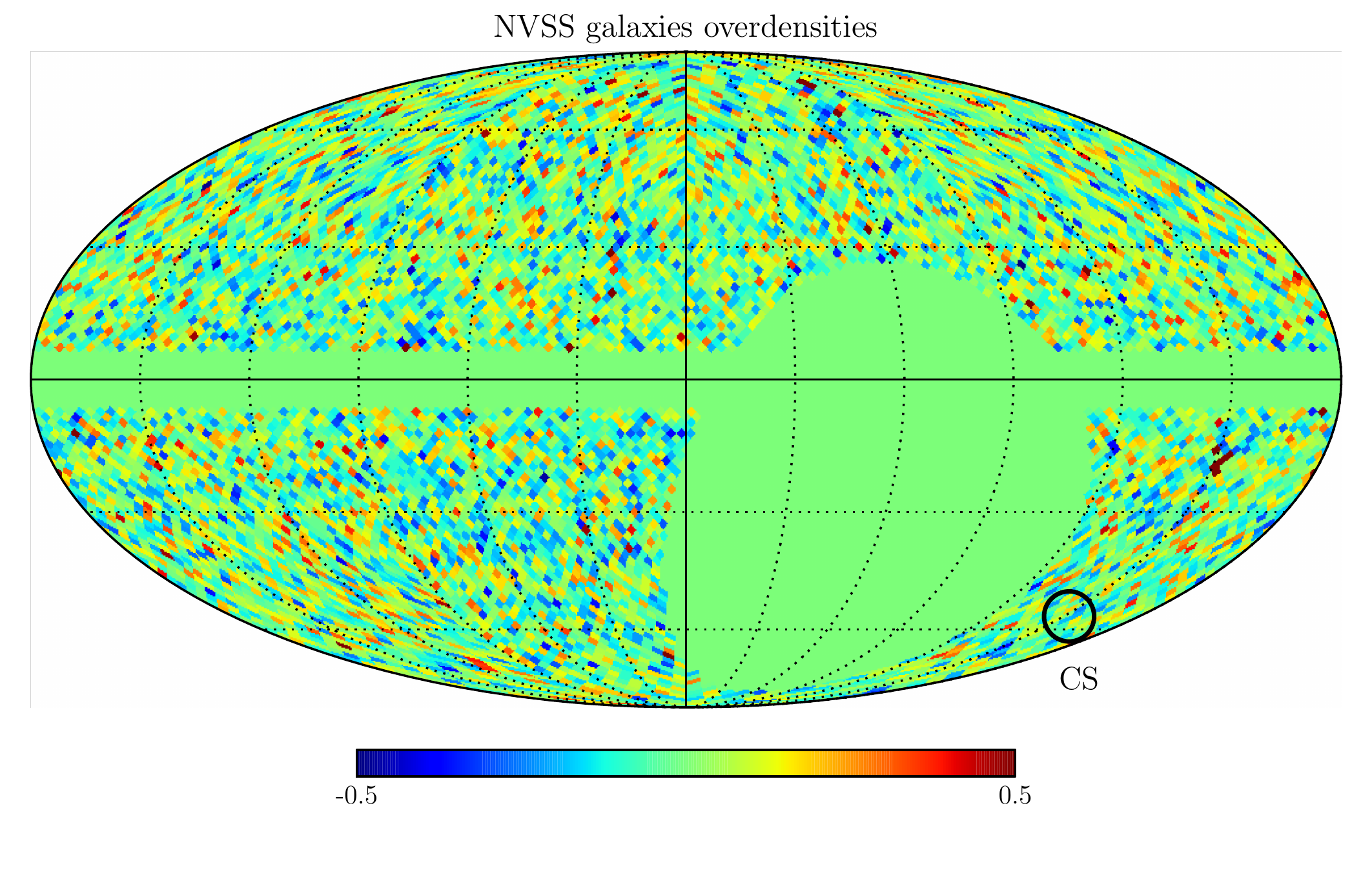}
  \caption{\footnotesize NVSS galaxies overdensities map, $N_{\text{side}}=32$. The cold spot location is showed in the map.}
  \label{fig:nvssmap}
\end{figure}
In particular we use data from the NRAO VLA Sky Survey (NVSS) \cite{NVSS}, a radio survey with enough sensitivity, redshift depth, and most importantly sky coverage to correlate with the ISW signal.

Summarizing its features briefly, this survey covers the northern hemisphere sky up to an equatorial latitude of $b=-40^{\circ}$ and detects approximately $1.4 \times 10^{6}$ sources above a flux threshold of 2.5 mJy.
Unfortunately the NVSS catalogue is known to be affected by a declination angle dependence in the mean number of observed galaxies due to the use of different radio telescope array configurations for observations at different angles.
To mitigate this problem, we apply a lower flux cut of 10 mJy at the catalogue taking advantage of the fact that more luminous sources are known to be less affected by this systematic (see \cite{marcos-caballero:2013}). This approach reduces our sample to $8\times10^{5}$ sources and a surface density of $\approx16~\text{deg}^{-2}$. 

At this point we use the RA and DEC positions of the remaining sources to create a pixelized map of the number of observed galaxies using HEALPIX\cite{gorski:2005}. Afterwards we mask a region of $\pm7^{\circ}$ from the galactic plane to avoid possible contamination from the Galaxy and the portion of the sky from the equatorial south pole to declination $-40^{\circ}$ which constitutes the blind spot of the survey. The final mask covers $28\%$ of the sky. 
Subsequently in each pixel we substitute the number of observed galaxies $n_{g}$ with their overdensities using the definition $\delta^{g}(\vec{x})=\frac{n_{g}(\vec{x})-\bar n}{\bar n}$. 
Finally we downgrade the map to $N_{side}=32$ obtaining the sky map showed in \reffig{nvssmap}.

However to model the correlation between galaxy overdensities and the other fields through \refeq{lineofsightint} we also need to know how these tracers are distributed as a function of redshift, \refeq{gal-kernel}.
It is very important to deal properly with this uncertainty because both the modeling of the ISW-galaxy cross correlation $C^{g\rm{ISW}}$ and the NVSS angular power spectrum itself used in our reconstruction algorithm are very sensitive to the redshift galaxy distributions. 
The NVSS survey does not measure the individual redshift of the sources and the optimal approach to model the redshift distribution of these radio sources have been a subject of debate in the literature.

Historically, early ISW analysis using the NVSS catalogue \cite{fosalba:2004}\cite{boughn:2004} modeled the redshift distribution using the radio luminosity function $\Phi(L,z)$ of Dunlop $\&$ Peacock \cite{1990MNRAS.247...19D}. 
More recently \cite{ho:2008} the NVSS redshift distribution has been obtained by cross-correlating the NVSS catalogue with other galaxy surveys with known redshift using a gamma distribution as template.  
Finally this problem has been approached taking advantage of the Combined-EIS-NVSS Survey of Radio Sources (CENSORS) \cite{brookes:2008}. This survey contained a sub-sample of 149 NVSS sources above 7.2 mJy within a 6 deg$^{2}$ patch in the ESO Imaging Survey (EIS) for which the redshift has been obtained spectroscopically or through the K-z relation. 
This is assumed as a fair sample of the redshift distribution and fitted firstly with a polynomial function \cite{de-zotti:2010} and recently with a gamma distribution. 
We refer the reader to \cite{marcos-caballero:2013} for a detailed analysis of NVSS properties and a comparison of different NVSS redshift distribution models using Bayesian evidence. 
Here we parametrize the redshift distribution using the best fit of \cite{marcos-caballero:2013}, a gamma distribution
\be\label{eqn:dndz-distr}
\frac{dn}{dz}=n_{0}\left(\frac{z}{z_{0}}\right)^{\alpha}\exp^{-\alpha z/z_{0}},
\ee
with $\alpha=0.36$ and $z_{0}=0.32$. The value of the parameter $n_{0}$ is chosen to normalize the total distribution to unity. The resulting distribution is shown in \reffig{Wfunction}.

As a final step, as a cosmological fiducial model, we chose the six parameters flat $\Lambda$CDM model that best fits the Planck+WP+lensing data combination \cite{planck-collaboration:2013}:
\begin{eqnarray}
&& \{ \Omega_ch^2, \Omega_bh^2, 100 \theta_*, \tau, n_s, 10^9 A_s \} \nonumber\\
&& \hspace{0.3cm} = \{ 0.118, 0.0223, 1.04167, 0.0947, 0.968, 2.215\}.\nonumber
\label{eqn:parameters}
\end{eqnarray}

\section{Tests and results}  \label{sec:results}

\subsection{Mock dataset reconstruction}\label{sec:mock}

We start by applying our reconstruction technique to a simulated dataset composed of maps of the lensing potential, CMB temperature and galaxy overdensity. Having knowledge of the input ISW map allows us to test our reconstruction procedure. We will show both how to interpret the results obtained with current data and possible improvements with future datasets.

To generate the mock datasets we first compute the theoretical auto and cross power spectra of the fields (ISW, CMB lensing and galaxies) from \refeq{cldef} with the exception of the primordial CMB temperature anisotropies obtained using CAMB. 
\begin{figure}[!htb]
\centering
  \includegraphics[scale=0.9]{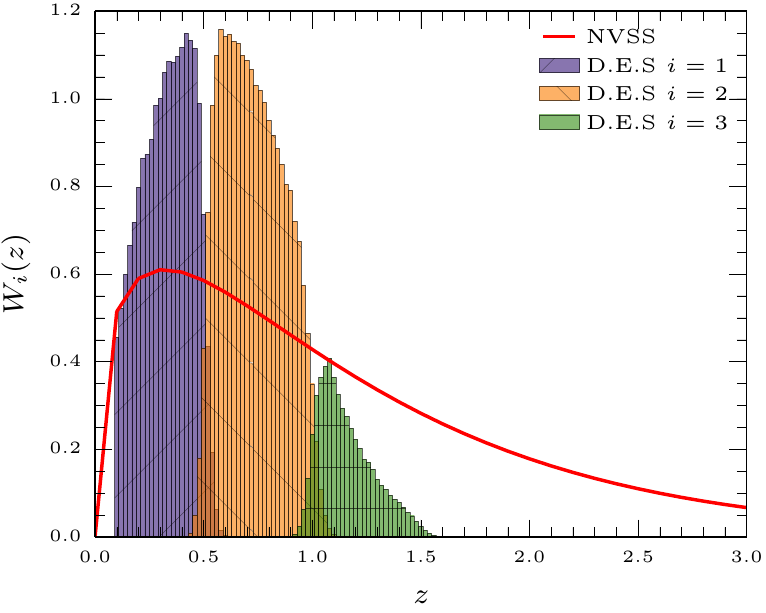}
  \caption{\footnotesize Example of window functions defined as in \refeq{zbin} and used to model the galaxies distribution. The real NVSS dataset (solid) is modeled following \refeq{dndz-distr} without any slicing applied. The mock data were created using the forecasted Dark Energy Survey (D.E.S) galaxies redshift distribution. To show how the redshift slicing in \refeq{zbin} works, the D.E.S redshift distribution is showed (using bars) after a convolution with three different slicing window functions. We do not use a top hat cut in redshift to take into account photometric uncertainties.}
\label{fig:Wfunction}
\end{figure}
A galaxy survey is characterized through its window function, Poisson noise and binning method.
We use the redshift distribution $dN/dz$ of the NVSS sample defined in \refeq{dndz-distr} and shown as the solid curve in \reffig{Wfunction}. No redshift binning is assumed because of the uncertainties in the NVSS redshift distribution.
We assume the galaxy bias $b_{i}$ is constant on the large scales of interest. Being constant, it can be determined using the auto-correlation of the galaxy map (assuming the underlying cosmology).

We use the computed power spectra to generate, for each field X, a realization of the associated spherical harmonics coefficients $a_{\ell m}^{X}$. They have been drawn from a gaussian distribution with mean zero and covariances $\langle a_{\ell m}^{X} a_{\ell' m'}^{Y}\rangle=\delta_{\ell \ell'} \delta_{m m'} C_{\ell}^{XY}$. The noise is generated in the form of $a_{\ell m}^{N}$ drawn from a Gaussian uncorrelated with other fields.
Armed with these components, we generate sky maps with a resolution of $N_{side}=32$, using HEALPIX.  

Finally we add to the primordial CMB the ISW map to mimic what we expect, on large scales, the observed CMB to look like. The observed CMB map, together with the noisy maps of CMB lensing and galaxy overdensities, properly correlated with each other, serve as our datasets. 

\begin{figure*}[htb]
\centering
  \subfloat[Input ISW.]{%
    \includegraphics[width=.24\textwidth]{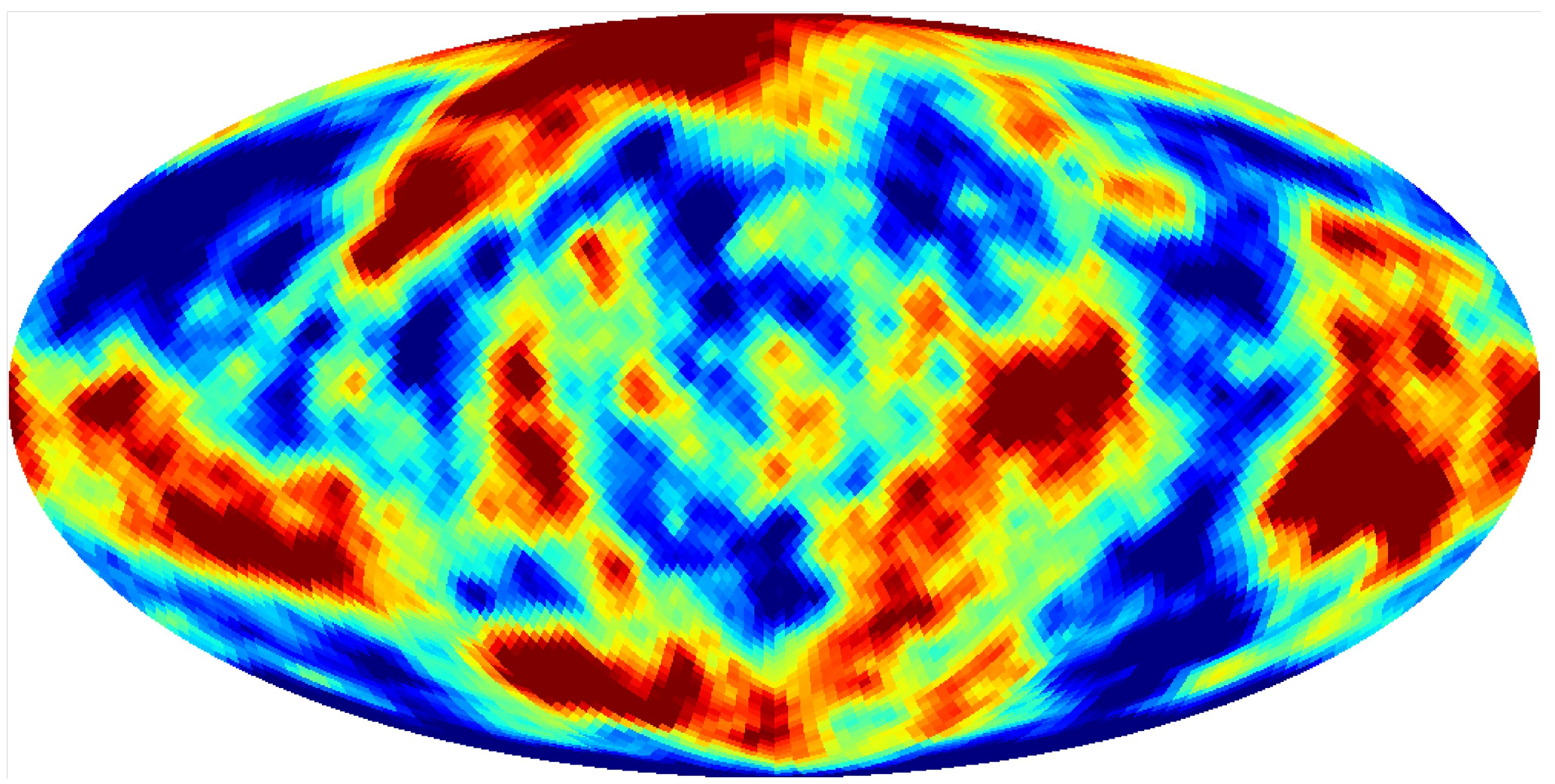}}\hfill
      \subfloat[\footnotesize Reconstructed ISW from NVSS + Planck Lensing + CMB.]{%
    \includegraphics[width=.24\textwidth]{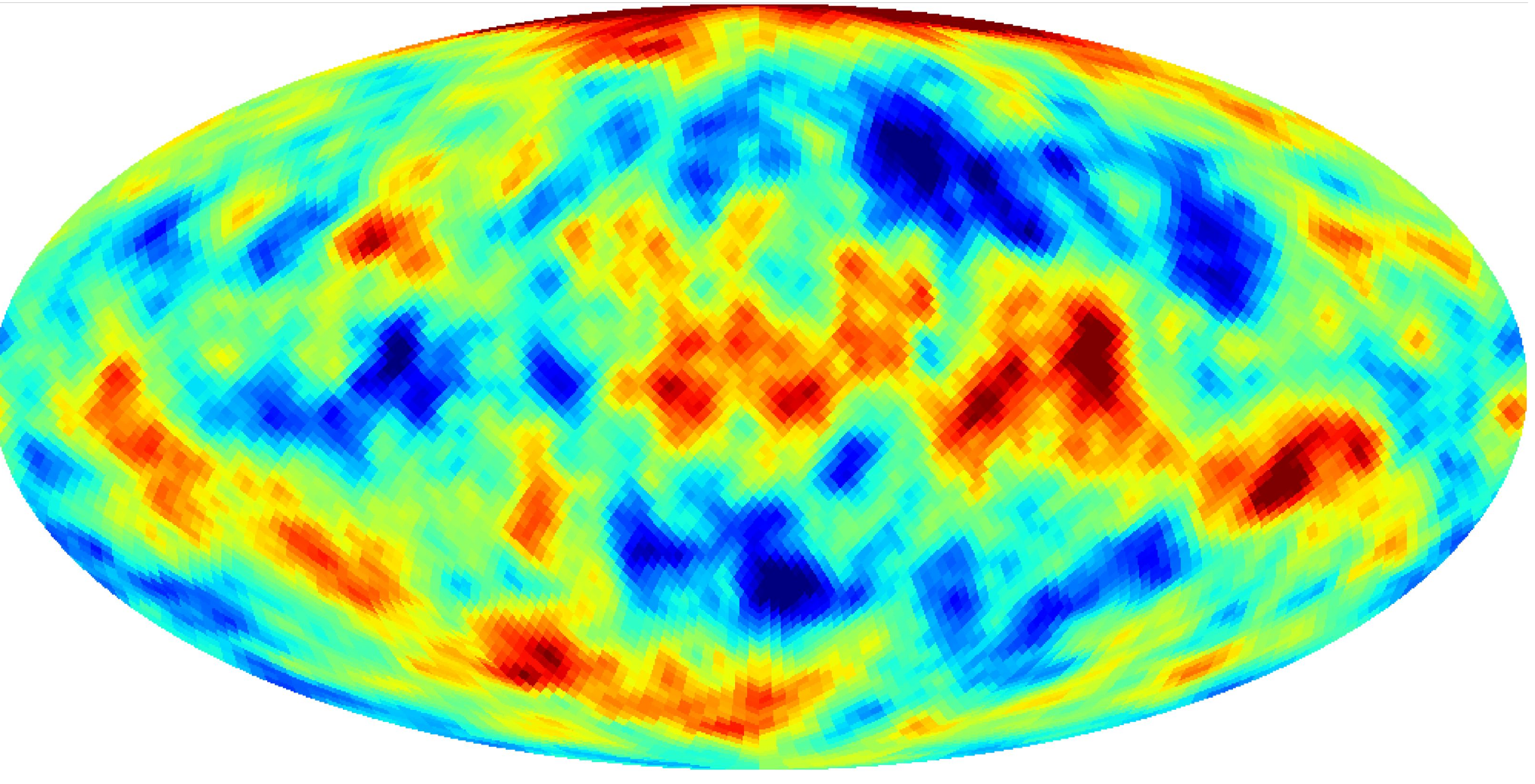}}\hfill
      \subfloat[\footnotesize Reconstructed ISW from Planck lensing.]{%
    \includegraphics[width=.24\textwidth]{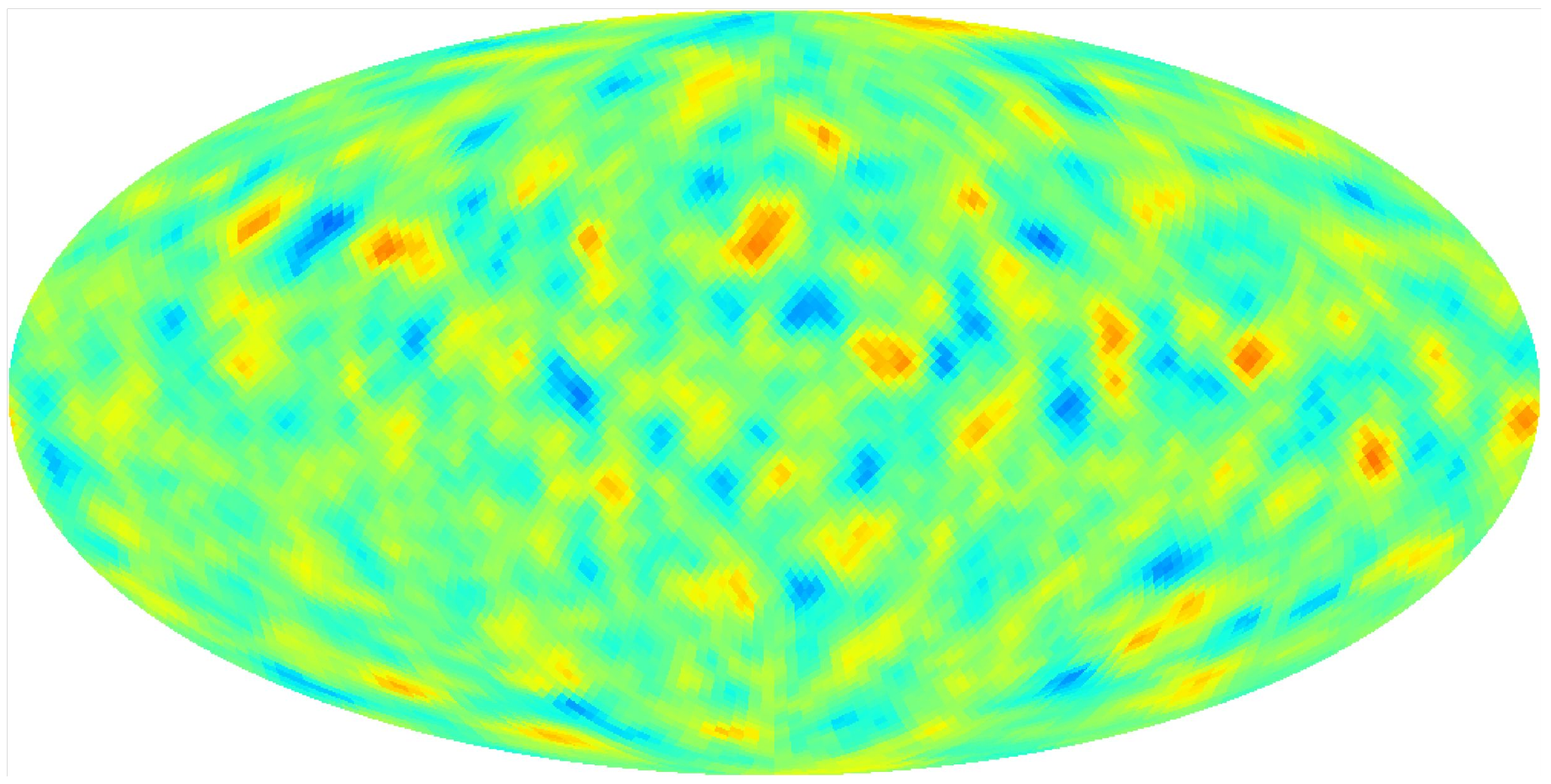}}\hfill
  \subfloat[\footnotesize Reconstructed ISW from a NVSS-like galaxy survey.]{%
    \includegraphics[width=.24\textwidth]{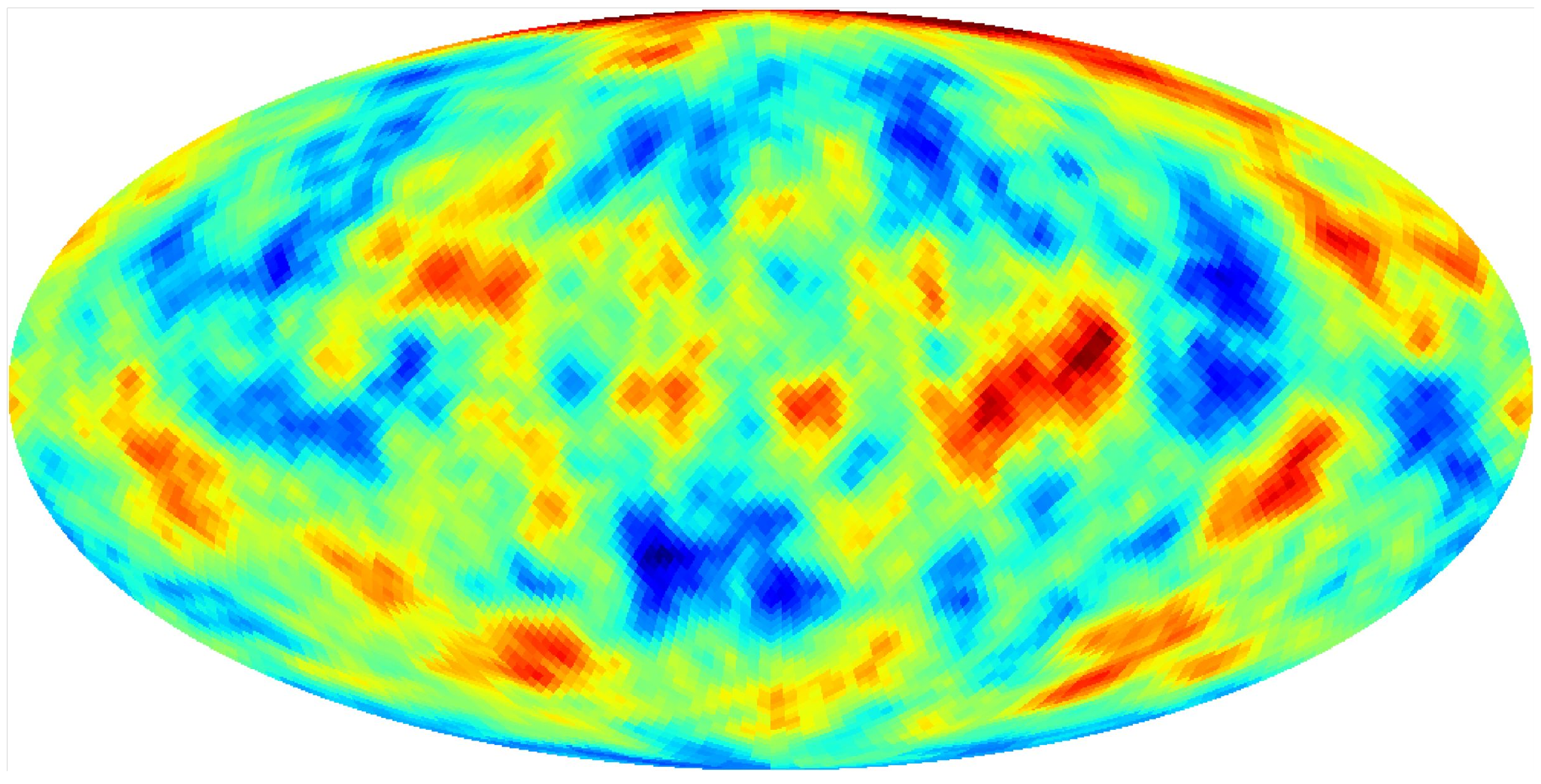}}\\
    \captionsetup[subfigure]{labelformat=empty}
     \subfloat[]{%
    \includegraphics[width=.34\textwidth]{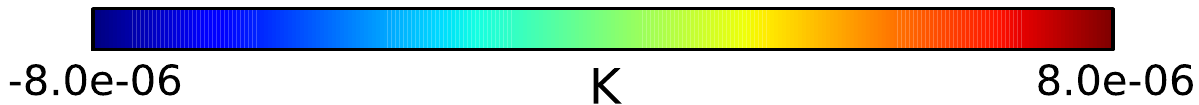}}\hfill
  \caption{\footnotesize Example of the reconstruction procedure applied to \textbf{realistic mock data} as described in \refsec{mock}. The simulated datasets have the same properties as the available real data: a Planck-like CMB temperature map, an NVSS-like galaxy distribution, and a lensing potential map without the low multipoles ($\ell<10$) and with the same reconstruction noise as Planck. The left panel show the input ISW map, and the other maps show the recovered ISW estimate using different datasets: (b) all 3 inputs; (c) Planck lensing only and (d) NVSS only. All the maps have a Healpix resolution $N_{\text{side}}=32$.
  }\label{fig:mockres}
\end{figure*}


An example of the reconstructed map from an NVSS-like survey, and a CMB lensing potential with the same reconstruction noise and the same multipoles ($\ell > 10$) as the one released by the Planck collaboration, and Planck temperature maps is shown in \reffig{mockres}. The reconstruction using all 3 datasets (Panel (b)) shares a number of the visual features apparent in the input map (Panel (a)). Panel (c) shows that very little of this information comes from the lensing potential maps, not surprisingly given the low-$\ell$ cut-off. Rather, most of the agreement stems from the information contained in NVSS, as is apparent in the agreement of the map reconstructed from NVSS only (d) and that from all 3 datasets (b).

\begin{figure}[!htb]
\centering
  \includegraphics[scale=1]{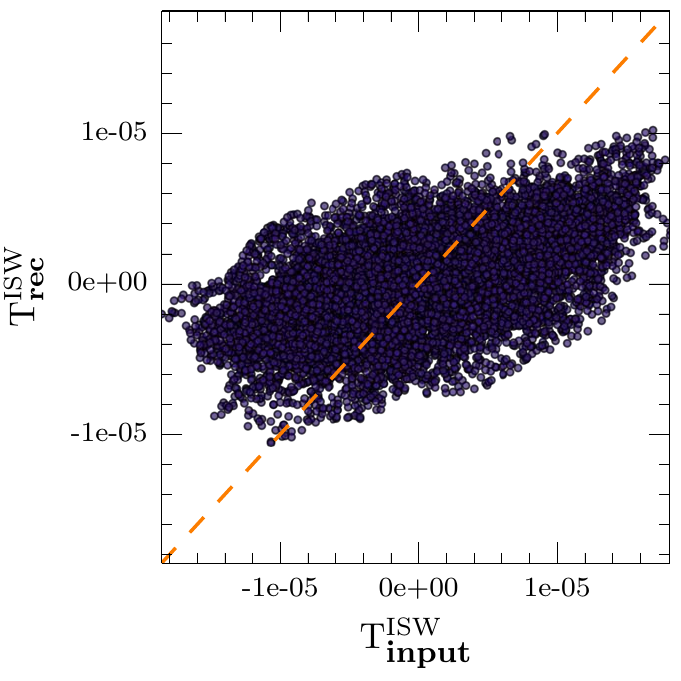}
  \caption{\footnotesize A pixel by pixel comparison between the input ISW map and the reconstructed one. The reconstruction here uses all 3 mock datasets (NVSS+temperature+lensing). Dashed line shows perfect correlation $\hat{T}^{\rm{ISW}}=\tisw$. For this analysis we use an Healpix resolution $N_{\text{side}}=32$ which corresponds to $12288$ pixels showed in the figure.}
   \label{fig:scatterT}
\end{figure}

Another way to inspect the quality of the reconstruction is to compare the ISW signal in the reconstructed map with the input map on a pixel by pixel basis. This scatter plot is shown in \reffig{scatterT}. If the reconstruction were perfect, the points would be scattered around the dashed curve for which $\hat{T}^{\rm{ISW}}=\tisw$. Although there is a clear trend of correlation, the overall agreement falls short, with the slope of the points smaller than unity.

\begin{center}
\begin{table} 
  {\begin{tabular}[t]{@{} |c|c| @{}}
    \hline 
    \multicolumn{2}{|c|}{Correlation coefficient $\rho$ from Mock Catalogs} \\
        \hline
   Galaxies 1 bin (NVSS) & 0.47  \\ 
   Galaxies 2 bin (DES) & 0.77  \\ 
   Galaxies 3 bin (DES) & 0.84  \\ 
   Lensing (Planck noise, all multipoles)& 0.42  \\ 
   Lensing (Planck noise, $\ell_{\text{min}}=10$) & 0.22  \\ 
   Lensing (half Planck noise, $\ell_{\text{min}}=10$) & 0.27  \\ 
    CMB Temperature & 0.39 \\ 
   ~~ 1bin (NVSS) + TT +Lensing ~~& 0.58  \\ 
    \hline
  \end{tabular}}
   \caption{\footnotesize  Defined in \refeq{rho}, $\rho$ quantifies the quality of the reconstruction. When not otherwise specified we use multipoles from $\ell_{\text{min}}=3 $ to $\ell_{\text{max}}=80$. For the galaxy survey, the redshift distribution is modeled following either the Dark Energy Survey (DES) with two or three redshift bins or the one-bin NVSS galaxy surveys. The lensing noise is the one released by Planck.}
\label{tab:rho}
\end{table}
\end{center}

A more quantitative way of estimating the quality of the recovered signal is to compute the correlation coefficient $\rho$ between the input and the reconstructed map, defined as:
\be \label{eqn:rho}
\rho=\frac{\langle \rm{ISW}~\tilde{\rm{ISW}} \rangle}{\sigma_{\rm{ISW}}~ \sigma_{\tilde{\rm{ISW}}}}
\ee
where \rm{ISW} ($\tilde{\rm{ISW}}$) is the input (reconstructed) ISW map and $\sigma$ is the variance of the map.
\reftab{rho} shows the value of $\rho$, averaged over 8000 simulations, for different datasets. The final row shows that, for these mock datasets, the mean value over all of simulations is $\rho=0.58$ when all the datasets are used together to reconstruct the signal. The realization depicted in \reffig{scatterT} had $\rho=0.56$, consistent with the visual sense that the slope was about half of its perfect value. The line labelled ``Lensing ...'' in \reftab{rho} shows that if only the lensing map is used the correlation is just slightly better ($\rho=0.42$ with all multipoles available) than temperature alone that leads to $\rho=0.39$. The galaxy map alone leads to $\rho=0.47$, not far from the the full information contained in all the maps. Clearly adding external maps, especially those that trace the galaxies, is a powerful way to extract more information about the ISW signal. The current lensing map is limited not only by noise but mostly by the absence of the low-$\ell$ multipoles. Notice that from our simulations (see line labelled ``Lensing ... $l_{\rm min}$ = 10'' in \reftab{rho}), we found that if we try to reconstruct the ISW signal using only lensing information, in the absence of the low multipoles $\ell<10$ in the lensing map, the correlation between the real ISW map and the reconstructed one is reduced by a factor of 2 ($\rho=0.22-0.27$), making it smaller than the one from temperature alone. Being able to reconstruct the lensing potential at this very large scales in the future will be crucial for this kind of science.

\begin{figure}[!htb]
\centering
  \includegraphics[scale=1]{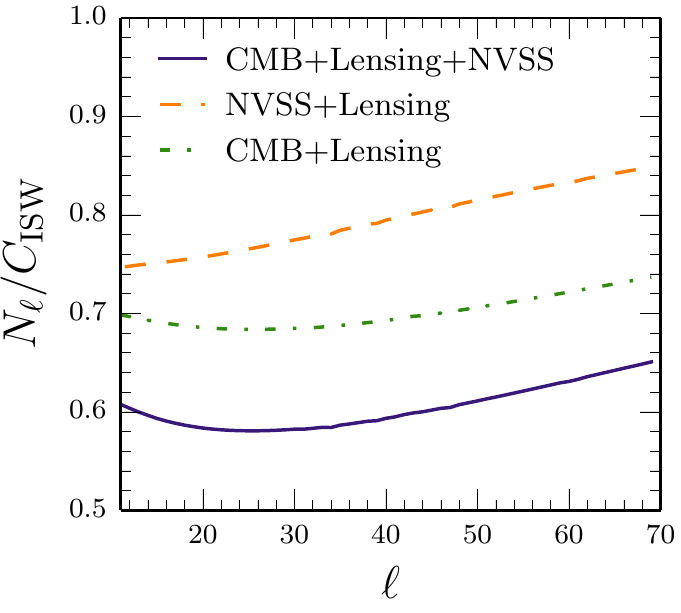}
  \caption{\footnotesize Reconstruction noise, defined in \refeq{error}, using different combinations of real datasets.}
  \label{fig:mockerrors}
\end{figure}

\vspace{1cm}

\subsection{Real Maps}\label{sec:real}

\begin{figure*}[htb]
\centering
      \subfloat[Combined.]{%
    \includegraphics[width=.48\textwidth]{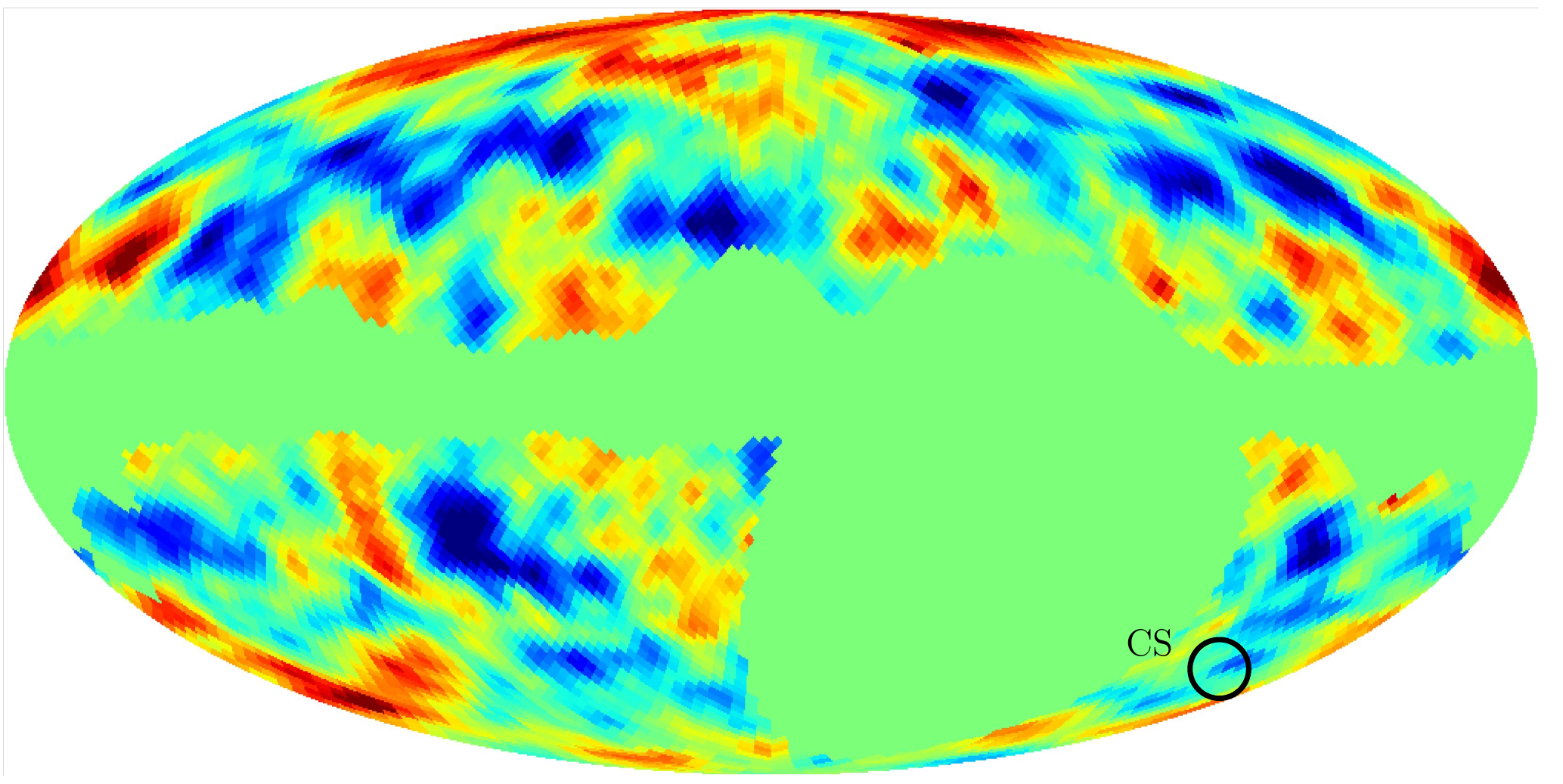}}\hfill
      \subfloat[Planck lensing.]{%
    \includegraphics[width=.48\textwidth]{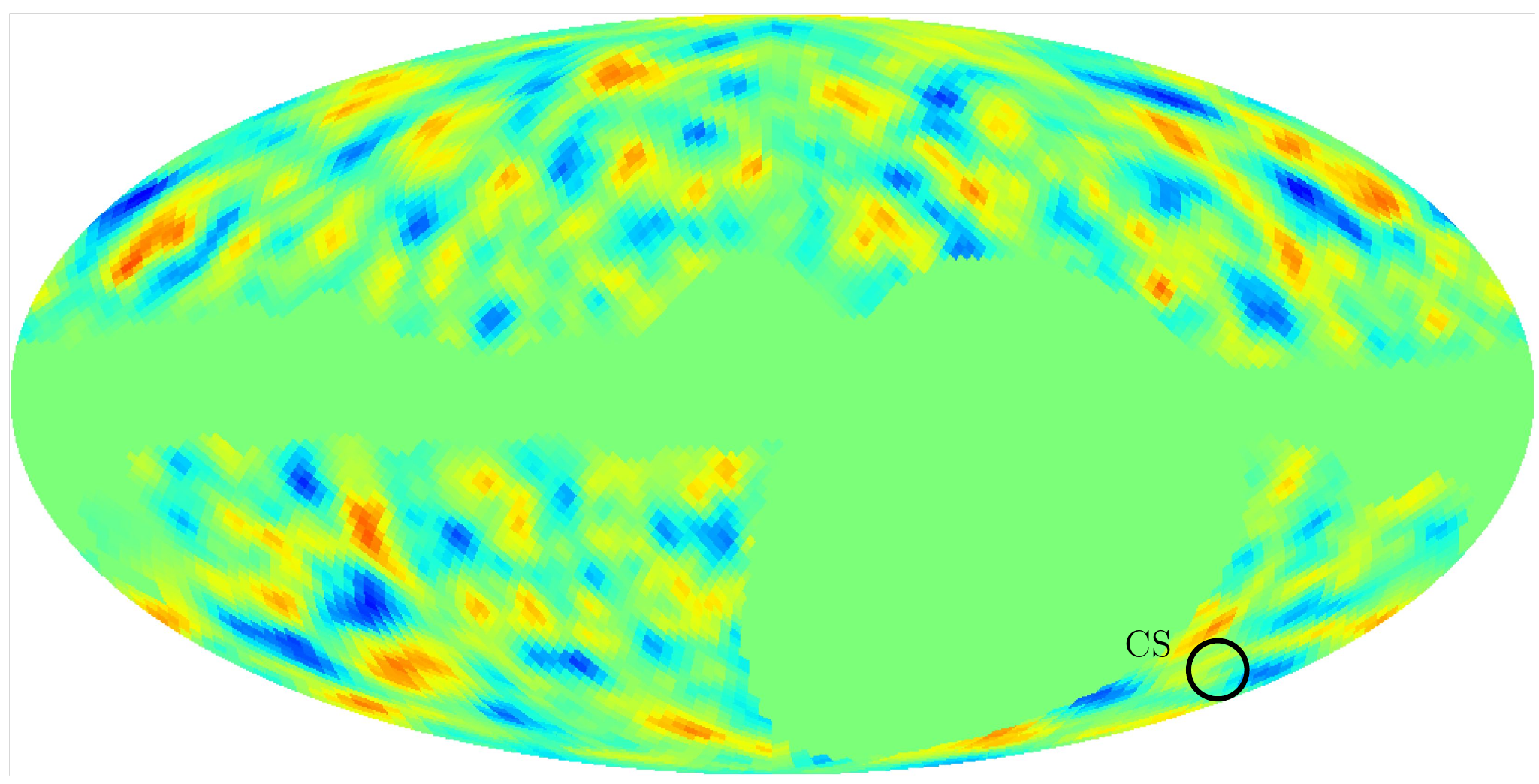}}\\[-2ex]
    \subfloat[Planck CMB Temperature.]{
        \includegraphics[width=.48\textwidth]{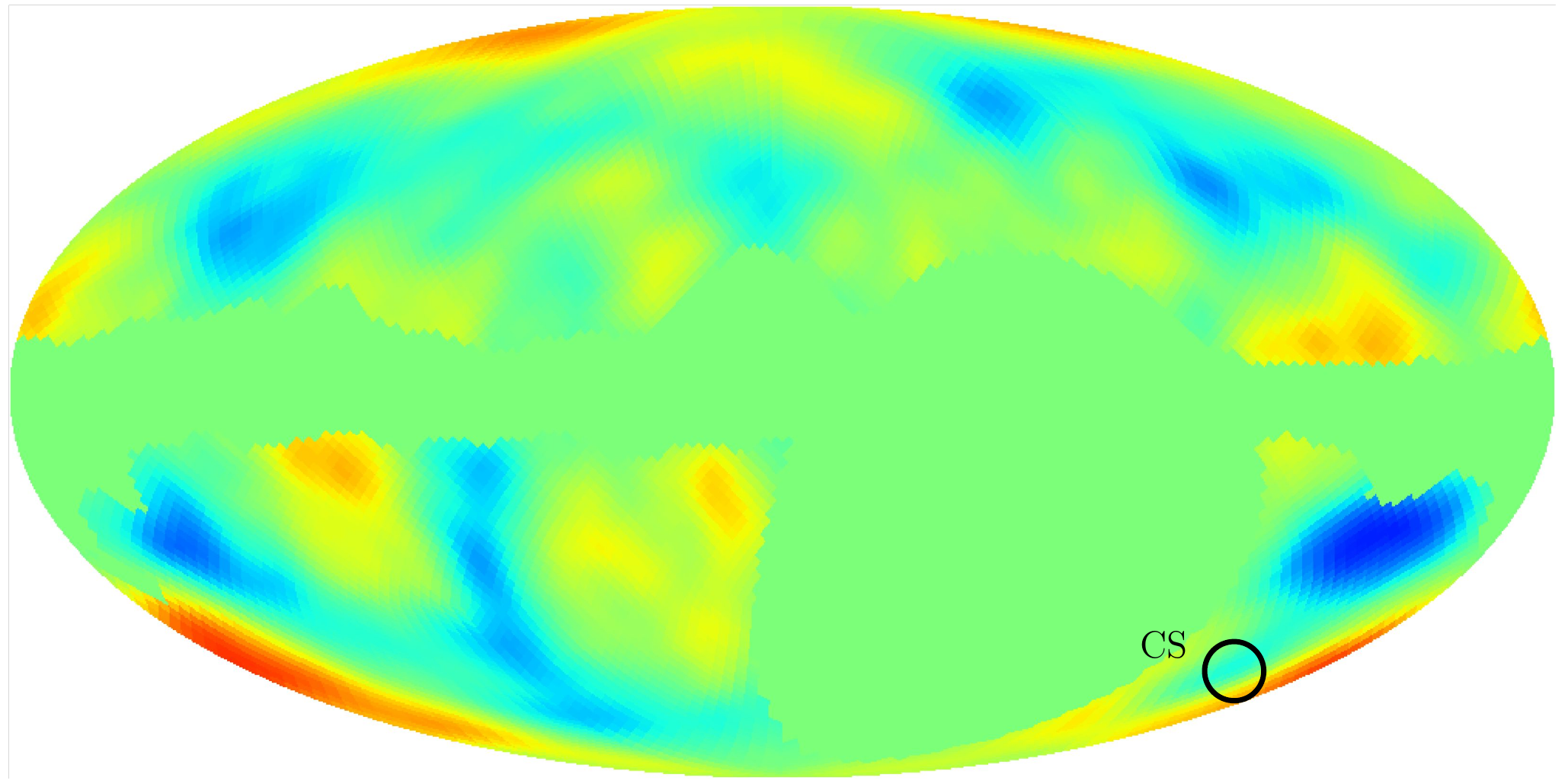}}\hfill
  \subfloat[NVSS Data.]{%
    \includegraphics[width=.48\textwidth]{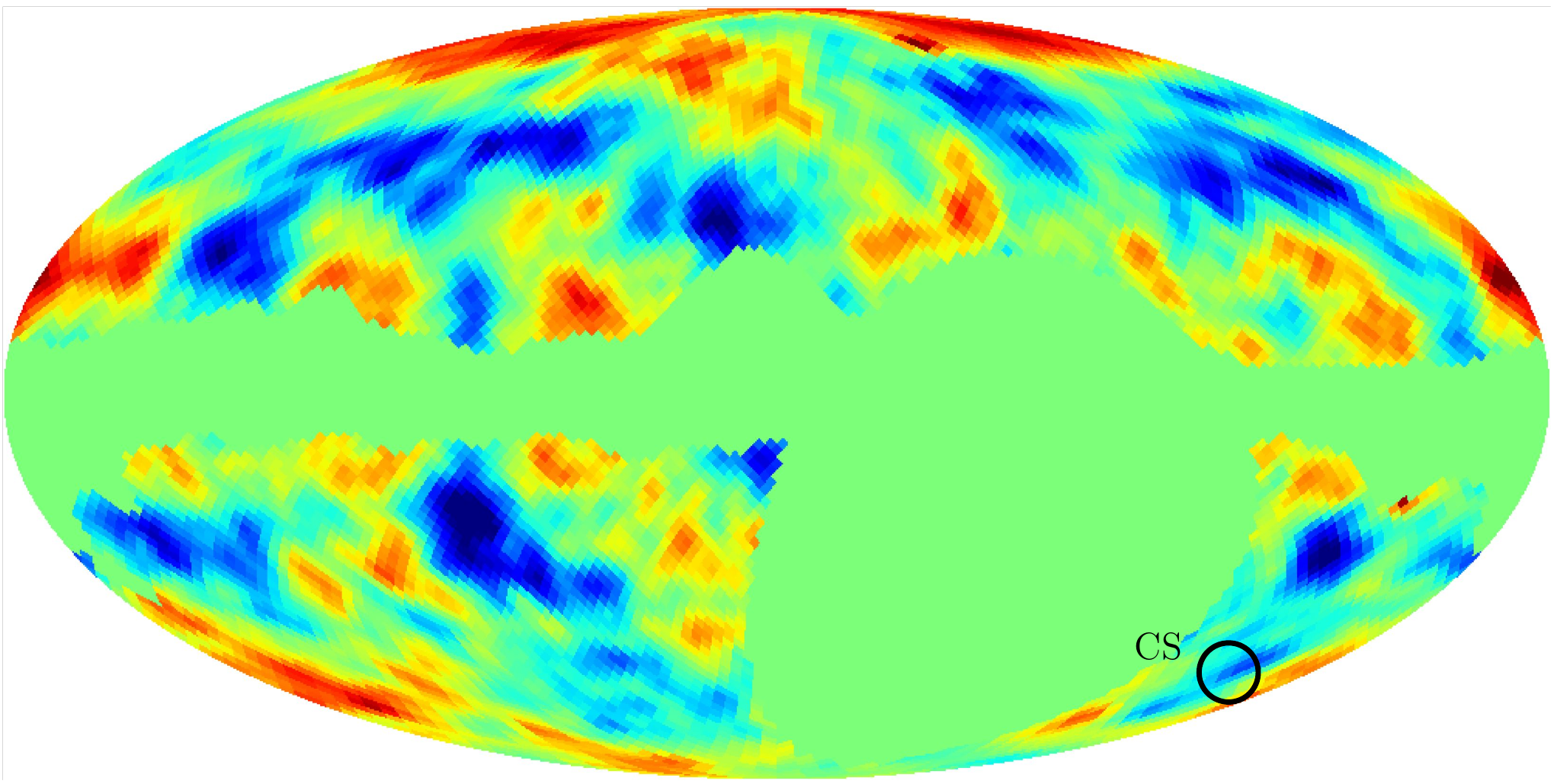}}\\
    \captionsetup[subfigure]{labelformat=empty}
     \subfloat[]{%
    \includegraphics[width=.34\textwidth]{./NVSS_bar.pdf}}\hfill
  \caption{\footnotesize Reconstructed ISW map from \textbf{real data} as described in \refsec{real}. Each figure shows the reconstructed map obtained using a different dataset. 
Fig. 8b uses only the gravitational lensing potential reconstructed by the Planck team. Fig. 8c is obtained from CMB temperature data only using a Wiener filter, Fig. 8d uses the galaxies overdensities from the NVSS catalogue. Our main results, Fig. 8a, optimally combined all the data.
The reconstruction uses multipoles from $\ell_{\text{min}}=3 $ to $\ell_{\text{max}}=80$ except for lensing potential information (Fig. 8b), available only for $\ell>10.$ The location of the Cold Spot is highlighted in the maps. The mask is the NVSS mask combined with the gravitational lensing one released by Planck. All the maps have a Healpix resolution $N_{\text{side}}=32$.}\label{fig:real} 
\end{figure*}


We now reconstruct an ISW map using real data. As described in \refsec{data} we use the NVSS radio galaxy survey data, the CMB temperature and the CMB reconstructed lensing potential data from Planck. The covariances contained in \refeq{Dmatrix} were computed from \refeq{cldef}. We assume that the galaxy bias is scale-independent and determine it as a free parameter that minimizes the chi-square between the NVSS power spectrum and the theoretical one. We obtain a value $b_{g}\simeq 1.96$ consistent with previous results \cite{ho:2008}.

The variance of the reconstruction for each multipole, as derived in \refeq{error}, is presented in \reffig{mockerrors} normalized to the prior given by $C^{\rm ISW}_{\ell}$. The top-most curve shows that the variance can be reduced by about 20\% if both NVSS and the lensing map are used; 30\% if the temperature and lensing maps are used; and 40\% if all three datasets are used. 

The reconstructed maps\footnote{The maps are available at \\ \url{http://astro.uchicago.edu/~manzotti/isw.html}} are shown in \reffig{real} and are in good agreement with previous literature \cite{barreiro:2013,planck-ISW}. 
As expected from simulations, the optimally reconstructed map (Panel a) is similar to the one constructed from NVSS (Panel d), as the galaxy map contains the most relevant information, especially given the low-$\ell$ cut-off in the lensing map. It is reassuring that the map constructed from the CMB temperature (Panel c) shares many of the large scale features evident in the map from NVSS. 

Also shown in \reffig{real} is the location of the Cold Spot in the CMB, a large scale anomaly with possible statistical significance \cite{cruz:2005,szapudi:2014,zhang:2010}. 
With these datasets, we estimate that the ISW component of the Cold Spot is significantly less than 10$\mu$K. This is an interesting piece of information that agrees with and complements earlier results~\cite{francis:2010,szapudi:2014,finelli:2014,2010MNRAS.403....2S}, who found that between $10-20\mu$K could be contributed by low redshift structures, particularly a void at $z\simeq 0.2$.  Given the redshift distribution shown in \reffig{Wfunction}, it seems unlikely that our analysis would pick up contributions from structures at such low redshifts. Therefore, our results strengthen the case that any late time imprints that might be responsible for the Cold Spot are likely to be at $z<0.3$.

\subsection{Projections: Future Surveys.}\label{sec:projection}
We conclude by studying how future surveys will help in reconstructing the ISW signal.

While the CMB temperature maps on large scales are unavoidably limited by cosmic variance and primordial signal, the CMB lensing reconstruction is limited by noise. This will be improved when data from future CMB polarization surveys are included in the lensing reconstruction \cite{hu:2002}. To account for this anticipated improvement, we have reduced the lensing reconstruction noise by a factor of two compared to the current Planck lensing map, and we assume that all multipoles will be available, not just $l>10$.

It is also evident that it will be very important to combine galaxy probes at different redshifts in the future. To make our projections concrete, we characterize the galaxy survey with the expected redshift distribution $dN/dz$ from the ongoing Dark Energy Survey. 
The procedure is exactly the same as described in \refsec{mock}, apart from the fact that we want to slice our survey in different bins.
We model the redshift binning process by assuming photometric redshift estimation gaussianly distributed around the true value with a rms fluctuation $\sigma(z)\simeq 0.05(1+z)$. In this case, a top hat cut in redshift results in a smooth overlapping distribution in actual redshift. 
Consequently $W_{i}(z)$, the selection function defining the i$^{th}$ slice, has a galaxy distribution \cite{hu:2004}:
\begin{eqnarray}
\label{eqn:zbin}
W_{i}(z) \propto {1\over 2} {dN(z) \over dz} 
\Bigl[{\,\rm erfc}\left({\Delta(i-1) -z \over \sigma(z)\sqrt{2}}\right) 
\\ \nonumber - {\,\rm erfc}
\left({\Delta i  - z \over \sigma(z)\sqrt{2}}\right)\Bigr],
\end{eqnarray}
where $\Delta$, the width of the bin, is chosen to cover properly the entire redshift range once the number of bins is given. 
Finally, we require that $\int W_{i}(z)dz = 1$. The window functions for 3 redshift bins in DES are shown in \reffig{Wfunction}.

\begin{figure*}[htb]
\centering
  \subfloat[Input ISW signal.]{%
    \includegraphics[width=.24\textwidth]{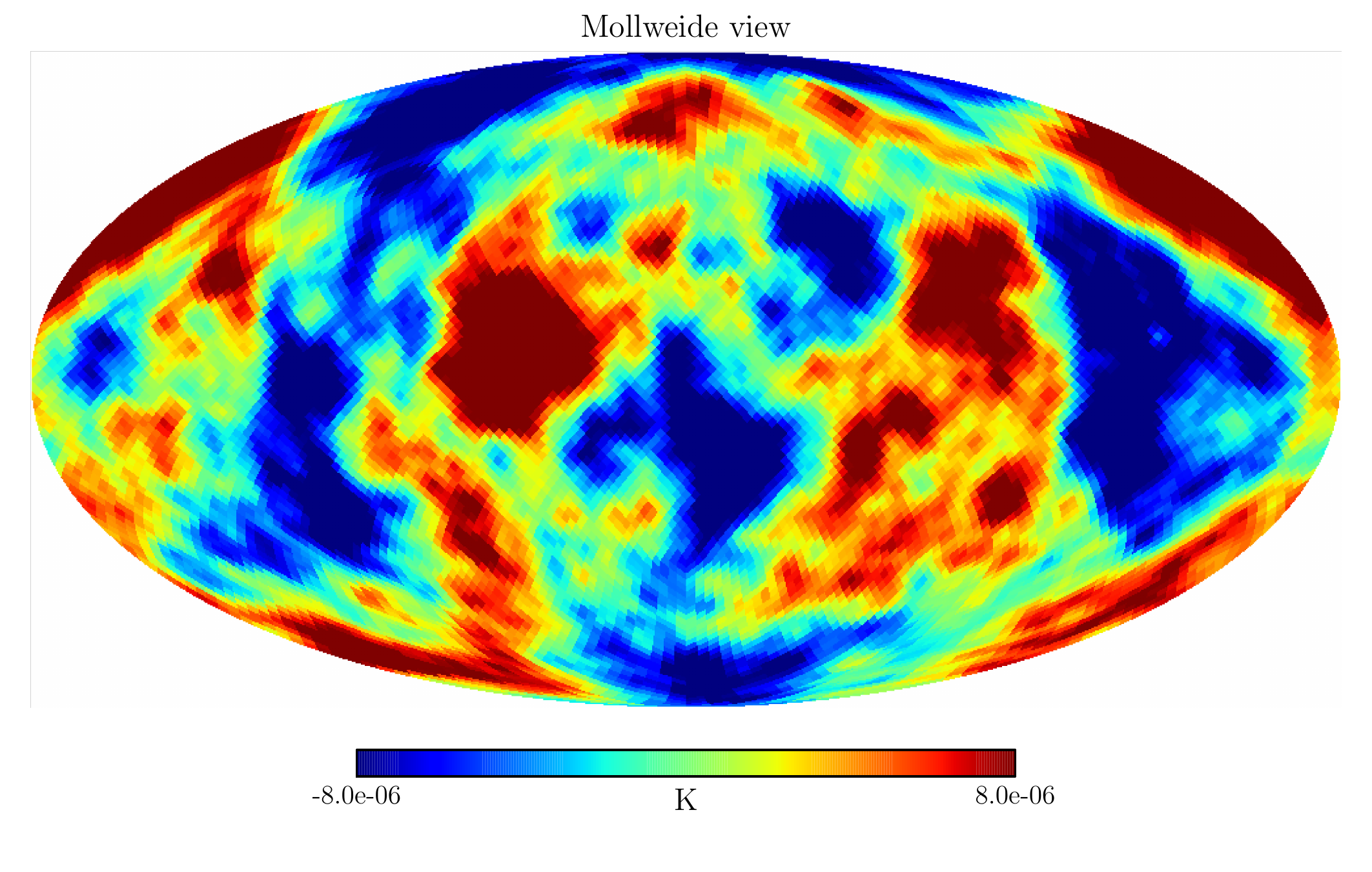}}\hfill
      \subfloat[Reconstructed ISW from Galaxies bins +Lensing+CMB Temperature.]{%
    \includegraphics[width=.24\textwidth]{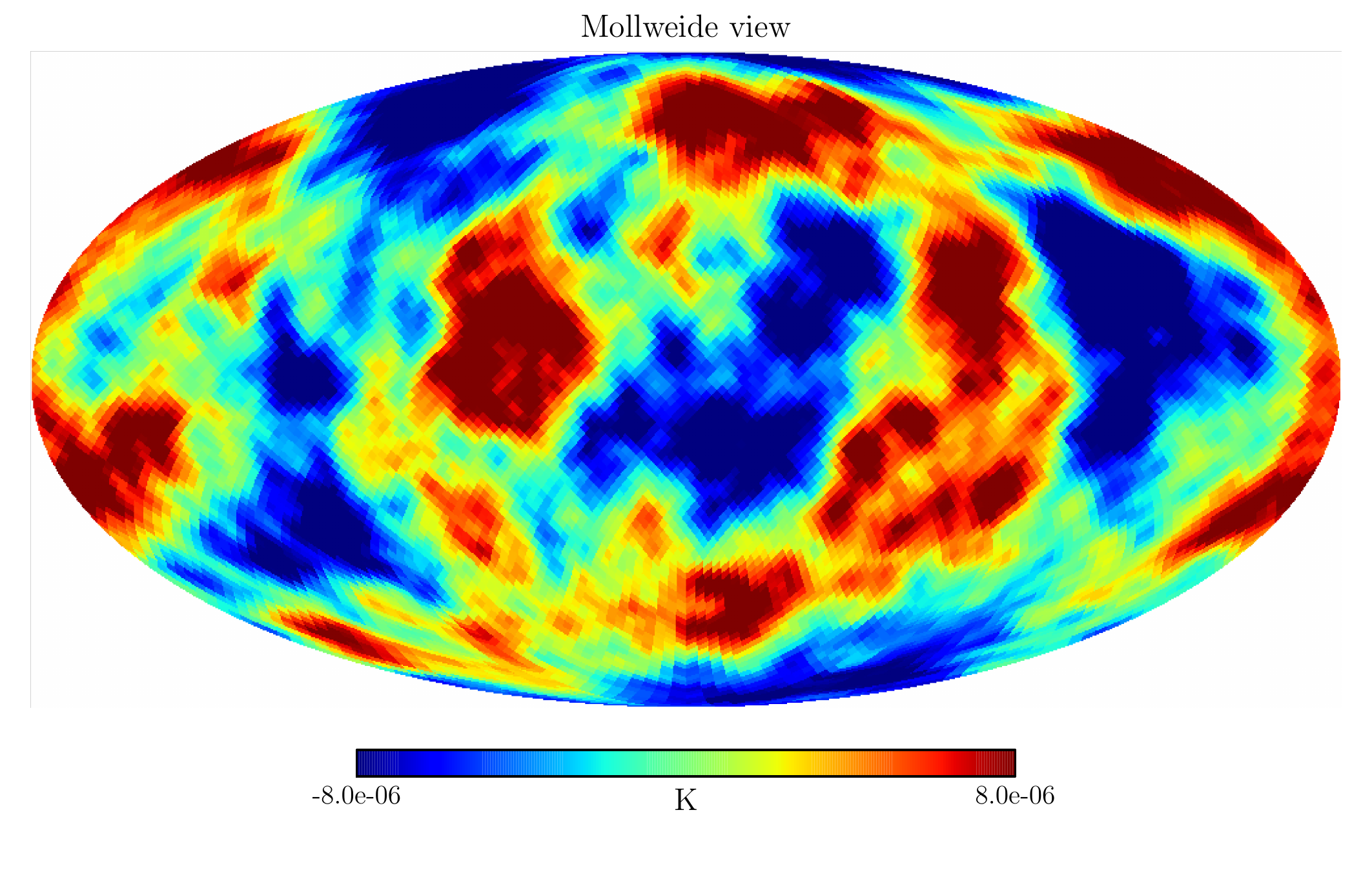}}\hfill
      \subfloat[Reconstructed ISW from Planck lensing with reduced reconstruction noise.]{%
    \includegraphics[width=.24\textwidth]{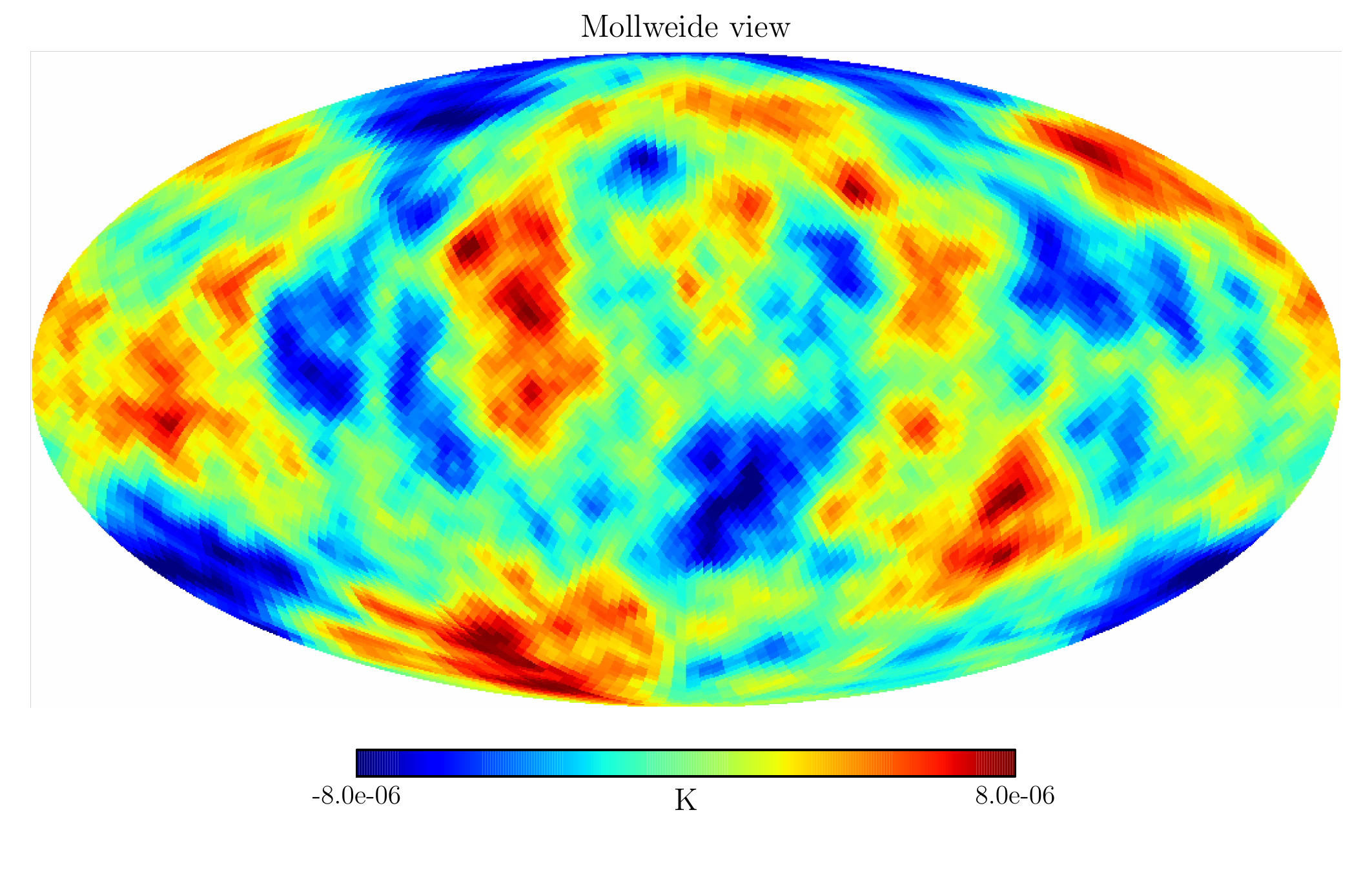}}\hfill
  \subfloat[Reconstructed ISW from Galaxies survey with 3 redshift bins.]{%
    \includegraphics[width=.24\textwidth]{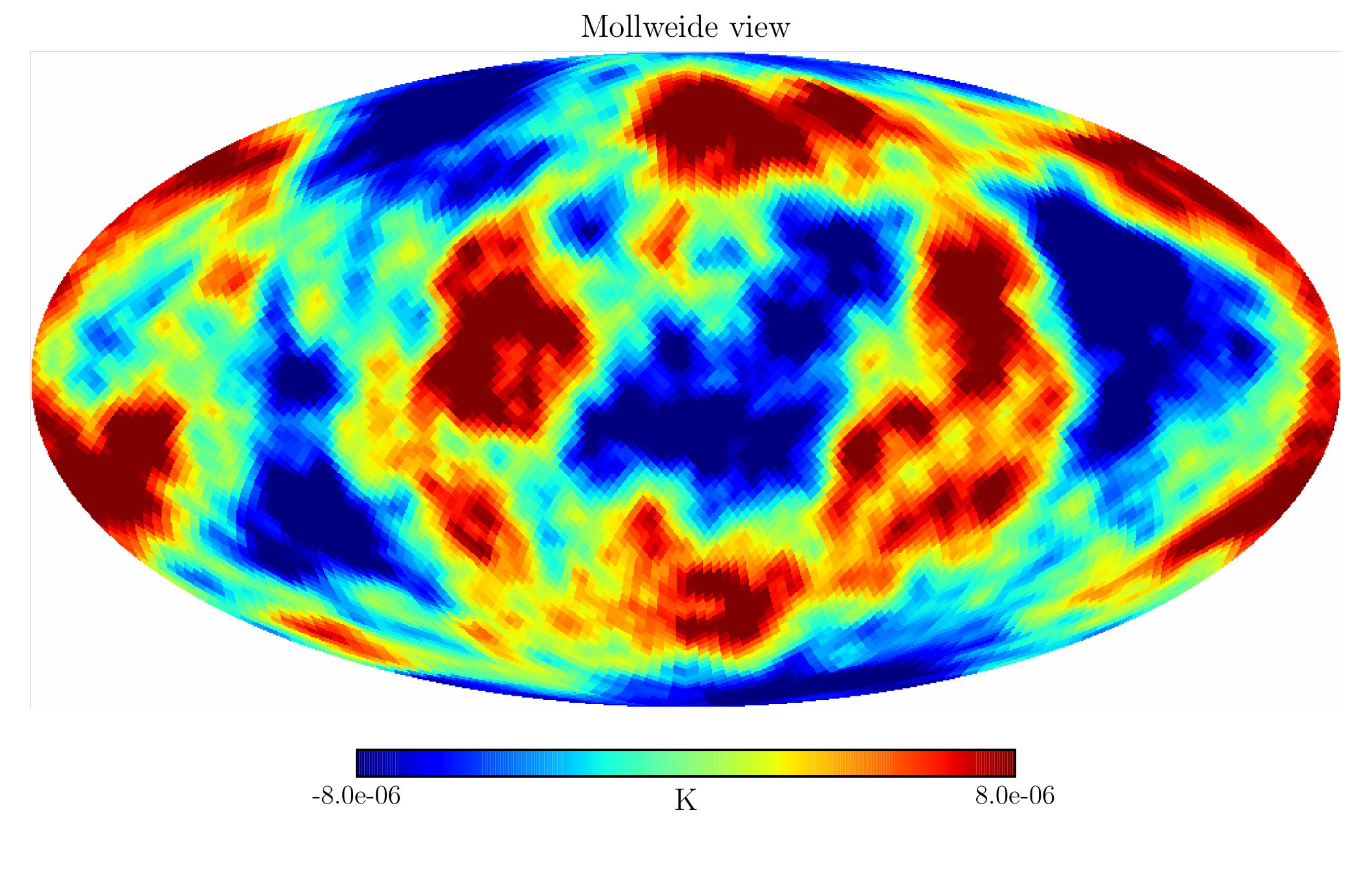}}\\
    \captionsetup[subfigure]{labelformat=empty}
     \subfloat[]{%
    \includegraphics[width=.34\textwidth]{./NVSS_bar.pdf}}\hfill
  \caption{\footnotesize Forecast ISW reconstruction with \textbf{future surveys}. We use our simulated data as described in \refsec{projection}. The reconstruction uses multipoles from $\ell_{\text{min}}=3 $ to $\ell_{\text{max}}=80.$ For the realization in the figure the signal to noise $\rho$ coefficients from the second to the left are $\rho=0.87,0.55,0.81.$}\label{fig:projres}
\end{figure*}

The reconstructed map from these datasets -- current temperature, improved lensing, and binned DES -- is shown in \reffig{projres}. 
It is clear that (see \reftab{rho}) slicing the galaxy survey improves the reconstruction, as long as there are enough galaxies in each redshift bin to keep the shot noise contribution of \refeq{ggnoise} small.

\section{Discussion}  \label{sec:conclusions}
The Integrated Sachs-Wolfe effect is the main secondary effect on the CMB temperature anisotropies at large scales. It is not only one of the most direct probes of dark energy properties but it can also be seen as a source of contamination for the primordial CMB anisotropies. For example it can be, at least partially, the cause of some of the large scale anomalies found in the CMB temperature data \cite{finelli:2014,rassat:2013,szapudi:2014}.

In this work we have investigated the reconstruction of the ISW signal map and its uncertainties using a likelihood technique able to combine different large scale structure tracers. In particular, we make use of the lensing potential reconstructed from its effect on CMB anisotropies and the distribution of galaxies in the NVSS survey. We test our technique on simulated datasets, comparing the true ISW map with the reconstructed one.

We show that, at the moment, galaxy surveys and the temperature map contain the most relevant information about the ISW signal. Including maps of the lensing potential does not significantly improve the reconstruction. This is not surprising since the current lensing maps are limited by noise and by the absence of the low multipoles ($\ell<10$), on precisely the scales where the correlation between ISW and lensing is largest.
However lensing will soon become powerful once the noise limitation will be alleviated by CMB polarization information. For example a reduction of the noise by a factor of two, which is a reasonable goal for the next release of Planck, will improve the reconstruction from lensing by almost 30\%. Moreover a better understanding of mean-field subtraction effects at low multipoles will extend to larger scales the available lensing information greatly enhancing the ISW reconstruction (see \reftab{rho}). 

Additionally future galaxy surveys, allowing us to use different redshift bins, will improve considerably the reconstructed signal.
In the future, as the datasets become larger, combining them optimally will be crucial to enhance the quality of the reconstruction. This is indeed one of the main motivation of this work.

Finally we have applied our technique to real data taking full advantage of the NVSS galaxy distribution, the Planck lensing potential and the Planck CMB temperature. The reconstructed ISW map is shown in \reffig{real}. Our technique was able both to recover results consistent with the literature \cite{barreiro:2013,planck-ISW} if limited to information coming from CMB lensing or galaxies distribution and to get a more accurate reconstruction when the entire dataset is used. 
We focus our attention to a possible secondary origin of the Cold spot in the CMB data. We find that the ISW component of the Cold Spot is significantly less than 10$\mu$K at the redshfits probed by our most powerful tracer, the NVSS survey. 
All the reconstructed maps are available for download at \url{http://astro.uchicago.edu/~manzotti/isw.html}.
\begin{acknowledgments}
We thank Eiichiro Komatsu for discussions that led to this idea and to him and Ryan Scranton and Eric Baxter for earlier unpublished work using it. This work was partially supported by the Kavli Institute for Cosmological Physics at the University of Chicago through grants NSF PHY-1125897 and an endowment from the Kavli Foundation and its founder Fred Kavli.
The work of SD is supported by the U.S. Department of Energy, including grant DE-FG02-95ER40896.
\end{acknowledgments}

\bibliographystyle{apsrev4-1}
\bibliography{./isw_paper}

\end{document}